\documentclass[%
 aip,
rsi,%
 amsmath,amssymb,
 reprint,%
]{revtex4-1}

\usepackage{graphicx}% Include figure files
\usepackage{dcolumn}% Align table columns on decimal point
\usepackage{bm}% bold math
\usepackage{scrextend}
\usepackage[margins]{trackchanges}
\usepackage{color}

\addeditor{VG}
\addeditor{HP}

\begin{document}

\preprint{AIP/123-QED}

\title[Linear and nonlinear optical probe of the ferroelectric-like phase transition in a polar metal, LiOsO$_3$]{Linear and nonlinear optical probe of the ferroelectric-like phase transition in a polar metal, LiOsO$_3$}

\author{Haricharan Padmanabhan}
 
\author{Yoonsang Park}
\affiliation{Department of Materials Science and Engineering, The Pennsylvania State University, University Park, PA 16801, USA}

\author{Danilo Puggioni}
\affiliation{Department of Materials Science and Engineering, Northwestern University, Evanston,
IL 60208, USA}

\author{Yakun Yuan}
\affiliation{Department of Materials Science and Engineering, The Pennsylvania State University, University Park, PA 16801, USA}

\author{Yanwei Cao}
\affiliation{Department of Physics and Astronomy, Rutgers University, Piscataway, NJ 08854, USA}

\author{Lev Gasparov}
\affiliation{Department of Physics, University of North Florida, 1 UNF Drive Jacksonville, FL 32224, USA}

\author{Youguo Shi}
\affiliation{Institute of Physics, Chinese Academy of Sciences, Beijing 100190, China}
 
\author{Jak Chakhalian}
\affiliation{Department of Physics and Astronomy, Rutgers University, Piscataway, NJ 08854, USA}

\author{James M. Rondinelli}
\affiliation{Department of Materials Science and Engineering, Northwestern University, Evanston,
IL 60208, USA}

\author{Venkatraman Gopalan}
 \email{vxg8@psu.edu}
\affiliation{Department of Materials Science and Engineering, The Pennsylvania State University, University Park, PA 16801, USA}

\date{\today}% It is always \today, today,

\begin{abstract}
LiOsO$_3$ is one of the first materials identified in recent literature as a 'polar metal', a class of materials that are simultaneously noncentrosymmetric and metallic. In this work, the linear and nonlinear optical susceptibility of LiOsO$_3$ is studied by means of ellipsometry and optical second harmonic generation (SHG). Strong optical birefringence is observed using spectroscopic ellipsometry. The nonlinear optical susceptibility extracted from SHG polarimetry reveals that the tensor components are of the same magnitude as in isostructural insulator LiNbO$_3$, except the component along the polar axis $d_{33}$, which is suppressed by an order of magnitude. Temperature-dependent SHG measurements in combination with Raman spectroscopy indicate a continuous order-disorder type polar phase transition at 140 K. Linear and nonlinear optical microscopy reveal 109$^o$/71$^o$ ferroelastic domain walls, like in other trigonal ferroelectrics. No 180$^o$ polar domain walls are observed to emerge across the phase transition.
\end{abstract}

\pacs{Valid PACS appear here}% PACS, the Physics and Astronomy
\maketitle

%Sample \add[HP]{added words} text to illustrate track changes feature. Sample text to illustrate track changes feature. Sample text to illustrate track changes feature. Sample text to illustrate track changes feature. Sample text to changed text illustrate track changes feature. Sample text to illustrate track changes feature. Sample \remove[HP]{text to} illustrate track changes feature. Sample text to illustrate track changes feature. Sample text to changed text illustrate track changes feature. Sample text to illustrate track changes feature. \note[HP]{This is a note.} Sample text to illustrate track changes feature. Sample text to illustrate track changes feature. Sample text to illustrate track changes feature. Sample text to illustrate track changes feature. Sample \change{text}{words} to illustrate track changes feature. Sample text to illustrate track changes feature. 

Polar metals are a relatively rare class of materials that exhibit ferroelectric-like long-range polar order in a metallic state \cite{Shi2013,Puggioni2014, Benedek2016}. These materials are of much interest because of the seeming incompatibility of polar order and metallicity, since free electrons in a metal are typically expected to screen the long-range electrostatic forces that stabilize polar ordering. The recent identification \cite{Shi2013} of a ferrolectric-like phase transition in metallic lithium osmate (LiOsO$_3$) has led to a flurry of activity \cite{Xiang2014, Giovannetti2014, Sim2014, Puggioni2014, LoVecchio2016, Benedek2016} to understand the origin of polar metallicity in these materials. From the point of view of applications, a ferroelectric-like structure in a metallic state leads to another intriguing question - how do functional properties typical of insulating ferroelectrics manifest themselves in a metal?

Optical properties are one such example, with the strong optical birefringence and nonlinear susceptibility of classical ferroelectrics such as LiNbO$_3$ and BaTiO$_3$ finding extensive application in optical materials and in optoelectronic devices \cite{Denev2011}. Nonlinear optical processes are also the foundation behind several important condensed matter phenomena such as higher harmonic generation \cite{Denev2011}, the optical Kerr effect, and photo-induced shift current \cite{Morimoto2016}. In this context, it is of interest to study and understand the optical properties of this unique class of polar materials. A recent study has reported \cite{Wu2016} that polar metals in the family of TaAs exhibit a giant nonlinear optical coefficient, with an enhanced response along the polar direction, providing further motivation for the present study. In this work, we consider one of the first identified \cite{Shi2013} polar metals, LiOsO$_3$. It is notable that this material has a lattice structure and polar phase transition that are isostructural to the ubiquitous insulating optical nonlinear materials LiNbO$_3$ and LiTaO$_3$ \cite{Shi2013}. A study of the optical properties of LiOsO$_3$ would hence be of interest not only from the point of view of potential applications, but also as a model system to understand the role of metallicity in the emergence of these properties. Furthermore, the nonlinear optical response can be used as a sensitive probe of important physical quantities such as spontaneous polarization \cite{Ramirez2009}, and phenomena such as phase transitions \cite{Harter2017,Ramirez2009}, ferroic domain formation \cite{Denev2011}, and coupling between different order parameters \cite{Ramirez2009}, all of which are critical to understanding the physics underlying this unique class of materials. Characterization methods based on nonlinear optics assume further importance given that the presence of free carriers makes it difficult to study these materials using traditional techniques like piezo-force microscopy and P-E loop measurements. 

\begin{figure*}
\includegraphics[width=0.8\linewidth]{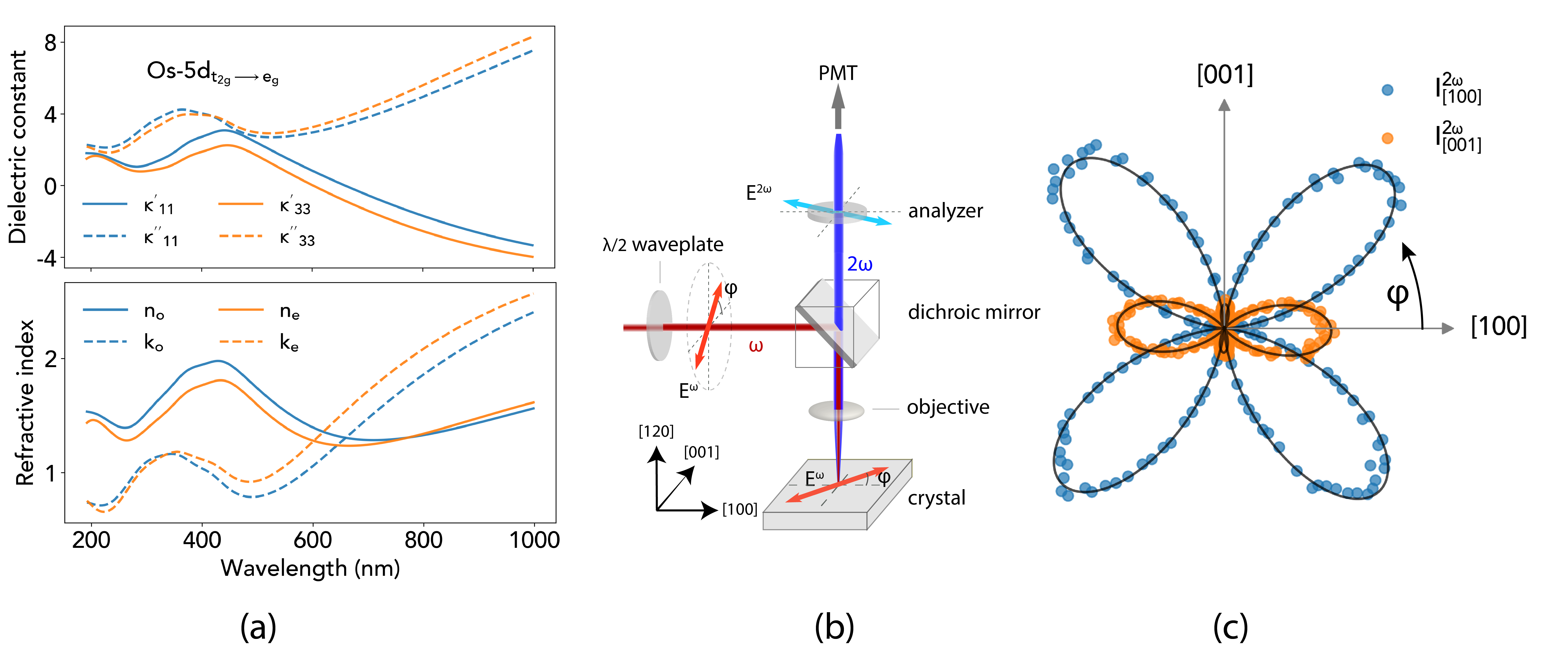}
\caption{\label{fig1} (a) Linear optical constants at 300 K obtained from spectroscopic ellipsometry. The top panel shows the dielectric constant, $\kappa = \kappa' + i\kappa''$, and the lower panel shows the refractive index $\tilde{n} = n + ik$. The parameters of the Lorentz oscillators used to fit the ellipsometry data can be found in Fig. S2. Figure (b) shows a schematic of the optical setup used for SHG polarimetry, and (c) the measured SHG polar plots at 20 K. The black lines are theory fits to 3m point group symmetry.}
\end{figure*}

In this work, the linear and nonlinear optical properties of single crystal LiOsO$_3$ were measured using spectroscopic ellipsometry and optical second harmonic generation (SHG) at a fundamental wavelength of 800 nm. Strong optical birefringence was observed in the complex linear refractive index. The optical SHG tensor coefficients extracted from SHG polarimetry revealed that the susceptibility along the polar direction is suppressed by an order of magnitude, compared to isostructural insulating ferroelectrics LiNbO$_3$ and LiTaO$_3$, as a consequence of the weak coupling mechanism \cite{Puggioni2014} that stabilizes the polar metal phase in LiOsO$_3$. We then use SHG in combination with Raman spectroscopy as a probe of the unique ferroelectric-like phase transition in LiOsO$_3$. These measurements showed that the polar phase transition in LiOsO$_3$ is continuous, exhibiting signatures of an order-disorder nature, with long-range order setting in at around 80 K. Linear and nonlinear optical microscopy and electron backscattering diffraction are used to identify 109$^o$/71$^o$ ferroelastic domain walls. No 180$^o$ polar domain walls were observed across the phase transition.

Optical second harmonic generation (SHG) is a second order nonlinear optical response to incident electromagnetic radiation. In this process, light incident on a material at frequency $\omega$ is converted into light at frequency $2\omega$, through the creation of a nonlinear polarization in the material, $P^{2\omega}_i = \epsilon_0\chi^{(2)}_{ijk}E^\omega_jE^\omega_k$, where $P^{2\omega}$ is the second harmonic polarization, $E^\omega$ is the electric field  of the incident light, and $\chi^{(2)}$ is the optical SHG susceptibility tensor. With $\chi^{(2)}$ being an odd ranked tensor, a non-zero SHG response is only allowed in noncentrosymmetric materials. SHG is thus a sensitive probe of inversion symmetry breaking, whether it is at interfaces such as surfaces, or in the bulk, through electric dipoles and multipoles. The bulk contribution is expected to dominate over the surface contribution \cite{Denev2011}. It is noteworthy that results of common theoretical models used to describe SHG, based on the anharmonic Lorentz oscillator, such as Miller's rule \cite{Boyd2003} and Kleinman symmetry \cite{Boyd2003,Dailey2004}, would in principle be invalid in polar metals, due to the finite intraband absorption associated with the lack of a bandgap.

%A priori, how are the nonlinear optical properties of a polar metal expected to vary from insulating polar materials? It is noteworthy that the results of common theoretical models used to describe SHG, based on the anharmonic Lorentz oscillator, such as Miller's rule \cite{Boyd} and Kleinman symmetry \cite{Boyd,Dailey2004}, would be invalid in polar metals, due to the finite intraband absorption associated with the lack of a bandgap.However, some general features can be expected of polar metals - for example, one would expect that the finite intraband absorption would result in a much larger SHG response than comparable insulating ferroelectrics, that are typically transparent to visible light. This has indeed found to be the case in TaAs in a previous study \cite{Wu2016}. Metals in general, also have a finite SHG response due to bulk multipole contribution from the free carriers. In a polar metal however, the contribution from bounded electric dipoles is expected to dominate over this. It will be seen in the following sections that screening by free carriers also possibly plays a role in the SHG response of polar metals.

LiOsO$_3$ single crystals were grown using solid state reaction under high pressure, as in the previous work by Shi et al \cite{Shi2013}. First, spectroscopic ellipsometry was used to obtain the linear optical constants at 300 K, between 200 nm and 1000 nm. The ellipsometry data was fitted using Lorentz oscillators (see Fig. S2), and the complex dielectric function $\tilde{\kappa}$ and refractive index $\tilde{n}$ were extracted. The complex dielectric function plotted in Fig. 1a shows a strong absorption at around 350 nm, which likely corresponds to Os-5d interband transitions from occupied $t_{2g}$ states to $e_g$ states above the Fermi energy\cite{Liu2015}. Notably, significant anisotropy is observed between the hexagonal [100] and [001] directions, contrary to previously reported data \cite{LoVecchio2016}. This anisotropy is likely primarily due to the Os-5d interband absorption, which was not accounted for in the reported work. The complex refractive indices in Fig. 1a are used in the subsequent analysis of the measured nonlinear optical constants.

The SHG tensor coefficients of LiOsO$_3$ were probed using SHG polarimetry as follows. The measurements were done in a far-field reflection geometry at normal incidence. The fundamental was a pulsed laser beam from a Ti:sapphire femtosecond laser with a wavelength of 800 nm (pulse width 80 fs, repetition rate 80 MHz). A beam with a power of 20 mW was focused on the (120) surface of a LiOsO$_3$ single crystal using a 50x objective, onto a spot size of $\sim$ 0.5 $\mu$m. SHG polarimetry was done following the schematic in Fig. 1b. The crystal was aligned so that the [120] axis was coincident with the fundamental, while the [100] and [001] were along mutually perpendicular directions in the plane of the probed surface. It was confirmed using electron back-scattering diffraction that the probed region was a single domain. The fundamental was linearly polarized, with the polarization within the (120) plane, making an angle $\phi$ with [100]. The reflected second harmonic light was collected using a photomultiplier tube, after passing through an analyzer. Note that all Miller indices are in the hexagonal setting.

Since LiOsO$_3$ is centrosymmetric (space group $R\bar{3}c$) at room temperature, the SHG at room temperature is likely primarily from inversion symmetry breaking at the surface and multipole contribution of bulk charges. The detector was calibrated so as to set this as the zero level. Measurements were then carried out at 20 K, which is below the reported \cite{Shi2013} non-polar ($R\bar{3}c$) to polar ($R3c$) phase transition temperature of 140 K. A pair of complementary polar plots were obtained by measuring the SHG signal as a function of the polarization of the fundamental, $\phi$, with the analyzer oriented along the [100] direction ($I^{2\omega}_{[100]}(\phi)$), and [001] direction ($I^{2\omega}_{[001]}(\phi)$), shown in Fig. 1c. To obtain the SHG tensor coefficients, these experimental polar plots were fitted to a theoretical model based on the $3m$ point group symmetry of LiOsO$_3$. The third rank tensor $\chi_{ijk}$ is usually written using the Voigt notation $d_{ij}$, in a pseudomatrix form. For $3m$ symmetry, the non-zero coefficients of $d_{ij}$ are $d_{15}=d_{24}$, $-d_{22}=d_{16}=d_{21}$, $d_{31}=d_{32}$, and $d_{33}$. Note that since LiOsO$_3$ is a metal with finite dispersion in the spectral range of interest, Kleinman symmetry, which allows for the permutation of certain tensor indices for materials that are dispersionless, is violated. The associated additional condition $d_{15}=d_{31}$ that is valid for isostructural insulators LiNbO$_3$ and LiTaO$_3$ is thus invalid for LiOsO$_3$. 

%The nonlinear polarization induced by the fundamental is then given by

\def\P{
\begin{bmatrix}
    P^{2\omega}_1 \\
    P^{2\omega}_2 \\
    P^{2\omega}_3
\end{bmatrix}}

\def\E{
\begin{bmatrix}
    E_1^2 \\
    E_2^2 \\
    E_3^2 \\
    2E_2E_3 \\
    2E_1E_3 \\
    2E_1E_2
\end{bmatrix}}

\def\dtensor{
\begin{bmatrix}
    d_{11} & d_{12} & d_{13} & d_{14} & d_{15} & d_{16} \\
    d_{21} & d_{22} & d_{23} & d_{24} & d_{25} & d_{26} \\
    d_{31} & d_{32} & d_{33} & d_{34} & d_{35} & d_{36} 
\end{bmatrix}}

\def\d3m{
\begin{bmatrix}
    0 & 0 & 0 & 0 & d_{15} & -d_{22} \\
    -d_{22} & d_{22} & 0 & d_{15} & 0 & 0 \\
    d_{31} & d_{31} & d_{33} & 0 & 0 & 0 
\end{bmatrix}}

%\begin{equation}
%\P \propto \d3m \E.
%\label{eq2}
%\end{equation}

Since the fundamental was linearly polarized in the (120) plane, $E_1=E_0cos(\phi)$, $E_2=0$, and $E_3=E_0sin(\phi)$. With the analyzer oriented along the [100] and [001] directions, the theoretical expressions for the two polar plots respectively simplify to 

\begin{equation}
\begin{split}
&I^{2\omega}_{[100]}(\phi)= |E^{2\omega}_{[100]}(\phi)|^2 \propto |d_{15}E_0^2sin^2(2\phi)|^2 \\
&I^{2\omega}_{[001]}(\phi) = |E^{2\omega}_{[001]}(\phi)|^2 \propto |d_{31}E_0^2cos^2(\phi)+d_{33}E_0^2sin^2(\phi)|^2.
\end{split}
\label{eq3}
\end{equation}

As Eq. \ref{eq3} shows, $d_{22}$ is not accessible using the (120) surface at normal incidence. This was instead obtained through an additional measurement at an angle of incidence of 45$^o$ (see Fig. S3). The experimental polar plots were fitted to these equations to obtain the ratios of the $d_{ij}$ coefficients. The SHG signal was calibrated with respect to congruently grown (001) LiTaO$_3$ to estimate the absolute magnitude of the coefficients. The expression derived by Bloembergen and Pershan \cite{Bor1962} was used to account for changes in reflectance due to differences in the linear refractive indices of LiTaO$_3$ \cite{Bruner2002} and LiOsO$_3$. The results are tabulated in Table 1, and the values of the SHG tensor coefficients of LiNbO$_3$ and LiTaO$_3$ are also given for comparison. 

Although $d_{15}$, $d_{22}$, and $d_{31}$ are of the same order of magnitude across all three materials, $d_{33}$ is lower by an order of magnitude in LiOsO$_3$. This behavior appears to be a consequence of the different coordination environment of Li and transition metal (M = Nb, Ta, Os) cations within the octahedra comprising the structures. In general the anharmonicity and associated optical nonlinearity in this family of materials are more sensitive to M-O acentric displacements than Li-O acentric displacements because of the larger nominal ionic charge on M ions compared to Li ions. LiOsO$_3$ exhibits substantially smaller M-O acentric displacements due to the weak coupling mechanism \cite{Puggioni2014} that stabilizes the polar metal state. This can be seen in the length of the long (l) and short (s) M-O bonds of LiOsO$_3$, as compared to those in LiNbO$_3$ and LiTaO$_3$, listed in Table S2. Qualitatively, the ratio of these bond lengths projected onto the [001] and [100] directions can be related to the degree of anharmonicity along these two directions, and hence to the SHG coefficients associated with them \cite{Tran2014,Cammarata2014,Cammarata2014a}. That is, M-O(l)/M-O(s) projected along [001] will affect $d_{33}$, whereas M-O(l)/M-O(s) projected along [100] will affect $d_{22}$ and $d_{31}$. Fig. 2 shows that the [001] projection of the acentric M-O displacements decreases significantly going from LiNbO$_3$ to LiOsO$_3$, which explains the suppressed $d_{33}$ in LiOsO$_3$. On the other hand, the [100] projection is relatively uniform (see Table S3), due to which $d_{22}$ and $d_{31}$ are each of the same order of magnitude across all three materials. A more in-depth analysis of this behavior is of interest, and will be the subject of future work.

\begin{table}
\caption{\label{tab:table1}SHG tensor coefficients of LiOsO$_3$ in pm/V, obtained from polar plots at 20 K, with those of LiNbO$_3$ and LiTaO$_3$ for comparison. Error bars are shown in parentheses.}
\begin{ruledtabular}
\begin{tabular}{cccc}
$d_{ij}$ & LiOsO$_3$\footnote{Values at 800 nm, 20 K from present work.} & LiNbO$_3$\footnote{\label{foot1}Values at 1060 nm, 300 K from reference  \onlinecite{Miller1992}.} & LiTaO$_3$\footref{foot1}\\
\hline
$d_{15}$ & $\pm$2.8 (0.4) & -5.2 (0.8) & -1.1 (0.2)\\
$d_{31}$ & $\mp$1.9 (0.2) & -5.2 (0.8) & -1.1 (0.2)\\
$d_{22}$ & $\mp$2.3 (0.3) & 2.8 (0.2) & 1.9 (0.2)\\
$d_{33}$ & $\pm$0.93 (0.1) & -36.3 (8.9) & -17.5 (2.2)\\
\end{tabular}
\end{ruledtabular}
\end{table}

\begin{figure}
\includegraphics[width=0.8\linewidth]{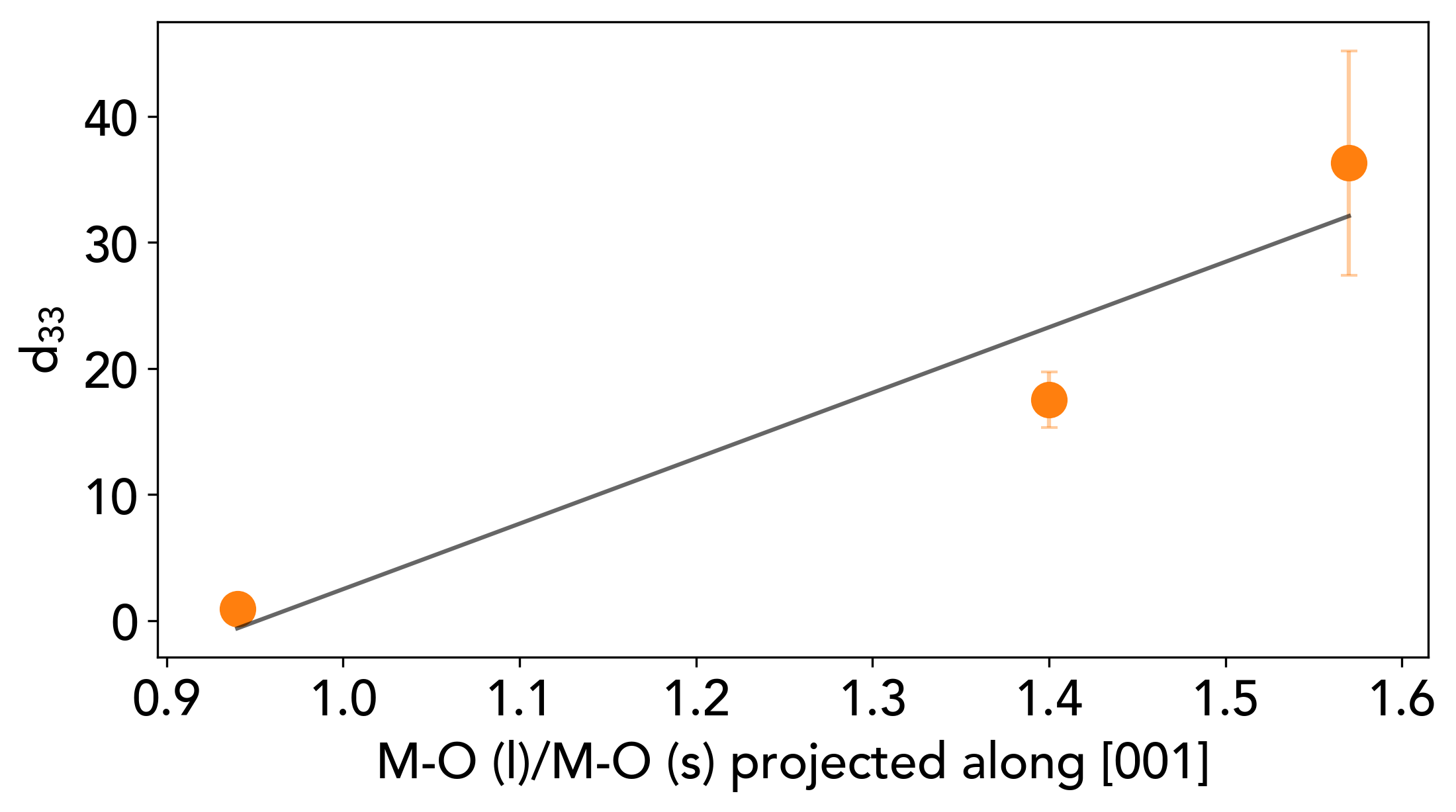}
\caption{\label{fig2}The variation in $d_{33}$ as a function of MO$_6$ acentric displacements along [001], as quantified by the ratio of the [001] projections of long (l) and short (s) M-O bonds in LiMO$_3$ with M = (Nb, Ta, Os). Bond lengths are taken from references \onlinecite{abrahams1966ferroelectric},  \onlinecite{abrahams1967ferroelectric}, and \onlinecite{Shi2013} respectively, and $d_{33}$ values from \onlinecite{Miller1992}.}
\end{figure}

The SHG response was also probed as a function of temperature to characterize the polar phase transition. The temperature detector was calibrated using a reference sample with a known phase transition. The absolute SHG signal as a function of temperature is plotted in Fig. 3, while the temperature dependence of the coefficients can be found in Fig. S4. The SHG response indicates a continuous polar phase transition with an onset near 135 K. The exact phase transition temperature is difficult to pinpoint because the signal is very close to the noise floor near the phase transition (see inset of Fig. 3). The fact that the signal is negligible above the phase transition temperature, and drastically increases as the temperature decreases below it indicates that the SHG response as probed in this geometry is strongly coupled to the dipoles in the bulk, and that this coupling dominates over other contributions. In fact the SHG response is coupled to the polar order parameter by a free energy term in the centrosymmetric phase (point group symmetry $\bar{3}m$) given by $F = -\chi_{ijkl}E_i^{2\omega}E_j^{\omega}E_k^{\omega}Q_l$, where $E^{\omega}$ refers to the fundamental electric field, $E^{2\omega}$ refers to the second harmonic electric field, and $Q$ refers to the order parameter corresponding to polar displacements. It can be shown \cite{Sa2000,Ramirez2009} that this term, which is zero above the phase transition temperature $T_0$, evolves into $F = -[\chi_{ijk3}Q_3]E_i^{2\omega}E_j^{\omega}E_k^{\omega}$, which is nothing but $-P^{2\omega}_iE^{2\omega}_i$, the energy due to interaction between second harmonic electric field and  polarization induced by a second order nonlinear optical susceptibility $\chi_{ijk}(T<T_0)=\chi_{ijk3}Q_3$. The SHG susceptibility $\chi_{ijk}$ is thus linearly proportional to the polar order parameter $Q$, and the measured SHG response $I^{2\omega}$ is proportional to $Q^2$. The temperature dependence of the SHG response in the present study is thus an indirect probe of the polar order parameter across the phase transition in LiOsO$_3$.

\begin{figure}
\includegraphics[width=0.9\linewidth]{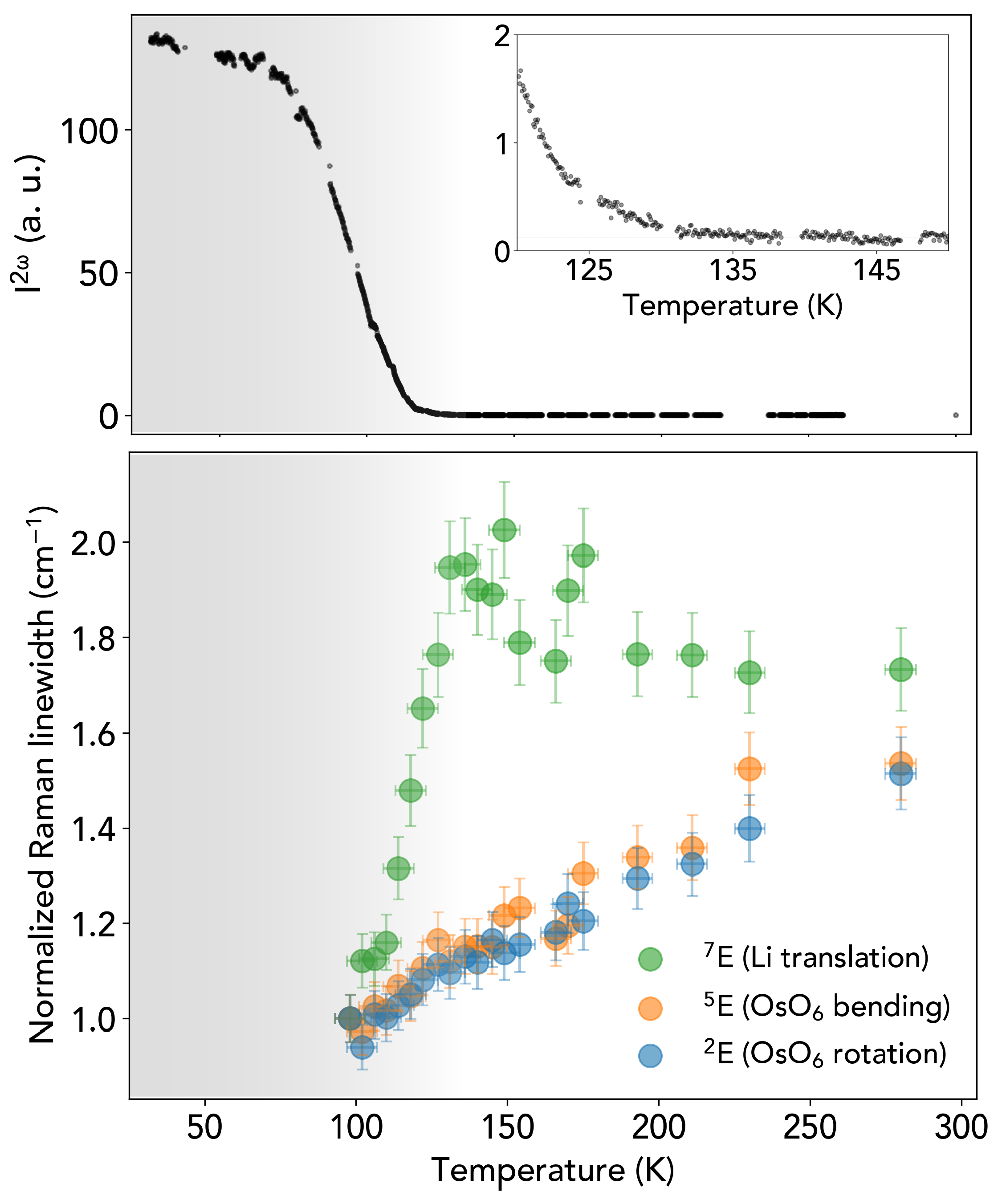}
\caption{\label{fig2}The temperature dependence of SHG is shown in the upper panel, with an enlarged plot of the dependence near the phase transition shown in the inset. The lower panel shows the Raman linewidths of three different phonon modes as a function of temperature, with $^2E$, $^5E$, and $^7E$ corresponding to modes at frequencies of 206 cm$^{-1}$, 402 cm$^{-1}$, and 492 cm$^{-1}$ respectively, at room temperature.}
\end{figure}

%Temperature-dependent Raman spectroscopy was used to study the behavior of phonon modes of LiOsO$_3$ across the phase transition. The peaks were assigned according to the previously reported work on Raman phonons in LiOsO$_3$ \cite{Jin2016}, and a full description of the measurements and the temperature-dependent spectra can be found in the supplementary material. For the R$\bar{3}$c to R3c phase transition considered in this work, one would expect Raman silent $E_u$ phonon modes in the  nonpolar phase to evolve into Raman active $E$ phonon modes in the polar phase. In this context, the Raman tensor for these phonon modes, given by $\big(\frac{\partial \alpha_{ij}}{\partial q}\big)_0$, where $\alpha$ is the polarizability, and $q$ is the normal mode coordinate of the considered phonon, is an additional order parameter connected with inversion symmetry breaking that emerges across the phase transition. The amplitude of the most prominent such mode, which corresponds to Os displacements in the (001) plane with a frequency of around 230 cm$^{-1}$, is plotted in Fig. 2a as a function of temperature along with the SHG response. The inset shows the Raman amplitude as a function of the SHG response. Recall that the SHG intensity is proportional to $Q^2$, and the Raman amplitude is proportional to $\big(\frac{\partial \alpha_{kl}}{\partial q}\big)_0^2$. The plot and the fit reveal that, $\big(\frac{\partial \alpha_{kl}}{\partial q}\big)_0 \propto Q^{\frac{1}{4}}$, approximately.

The SHG response and the associated polar order parameter gradually increase as the sample is cooled below 140 K, before increasing rapidly and saturating at around 80 K, as seen in Fig. 3a. It is pertinent to ask whether this behavior is due to the occurrence of an order-disorder phase transition as opposed to a displacive phase transition, as this is a point of contention in the literature \cite{Shi2013, Sim2014, Liu2015, Jin2016}. To shed more light on this, Raman spectroscopy measurements were carried out to study the behavior of phonon modes across the phase transition. Temperature increase due to laser heating was taken into account using the Stokes-anti-Stokes relationship. To begin with, no Raman active soft modes that might result in a displacive phase transition were observed in the Raman spectra, consistent with previous studies \cite{Jin2016}. Raman linewidths are directly related to disorder in the system, and hence when studied as a function of temperature, can be used to identify order-disorder phase transitions, as done in previous work on complex oxides \cite{Singh2008,Singh2014}. Previous studies showed a large decrease in Raman peak linewidths below 140 K, and attributed that to the order-disorder nature of the phase transition \cite{Jin2016}. In the present work, we repeat these measurements and expand on the analysis, paying particular attention to the Raman modes corresponding to displacements of Li atoms.  The proposed order-disorder transition consists of ordering of off-centered Li atoms, so it is reasonable to expect that if the phase transition is indeed order-disorder in nature, the Raman phonon modes consisting of Li displacements would exhibit the largest change in linewidth. The linewidths of the three dominant $E$ modes (see Fig. S5 for peak labels) are plotted in Fig. 3b, of which only one, $^7E$, involves the displacement of Li atoms. Clearly this mode, consisting of antiparallel Li translation perpendicular to the polar direction, shows the largest increase in linewidth, with a clear discontinuity in the slope near the critical temperature. This linewidth increase is more than twice that of the $^2E$ and $^5E$ modes, which consist of OsO$_6$ octahedra bending and rotation respectively. Additionally, the present measurements were limited to around 98 K, but the Raman peak linewidths in the work by Jin et al\cite{Jin2016}, with data collected down to 10 K, appear to saturate at the same temperature as the SHG response in the present work, at around 80 K, providing further evidence that these two results may be complementary signatures of the onset of long-range polar order through an order-disorder transition. 

\begin{figure}
\includegraphics[width=0.8\linewidth]{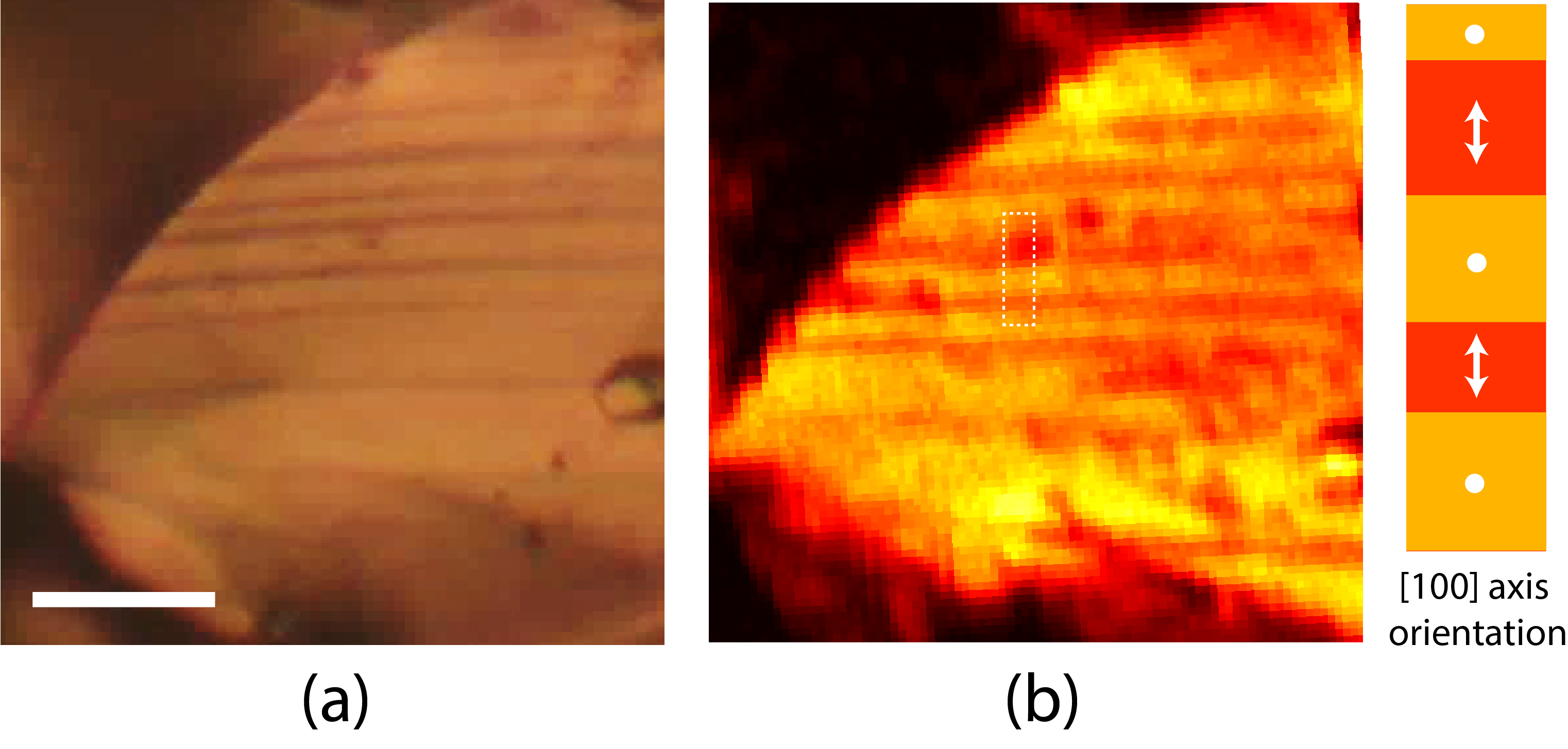}
\caption{\label{fig3}Ferroelastic domains in LiOsO$_3$ with (a) showing a linear optical micrograph at 300 K, and (b) a nonlinear optical micrograph using SHG reflectance at 20 K. A schematic of the domains is shown on the right, with the arrows denoting the direction of the [100] crystallographic axis, obtained from electron back-scattering diffraction (see Fig. S5). The scale bar in (a) is 10 $\mu m$.}
\end{figure}

Finally, linear and nonlinear optical microscopy was used to study the influence of the ferroelectric-like phase transition on the mesoscale structure of LiOsO$_3$. The linear optical micrograph in Fig. 4a shows a striped domain structure observed in the nonpolar phase. These stripes were oriented using electron back-scattering diffraction (EBSD) (see Fig. S6) and identified as ferroelastic domains, with the [100] crystallographic axes in adjacent domains forming an angle of $90^o$ relative to each other, as shown in Fig. 4b. This geometry is consistent with 71$^o$/109$^o$ polar domain walls formed by the breaking of four-fold symmetry in the lattice, going from a high temperature cubic $Pm\bar{3}m$ structure to the room temperature trigonal $R\bar{3}c$ structure. The sample surface was also studied using scanning SHG microscopy at 20 K. The SHG map confirms that these domains are polar below the phase transition, and the contrast between stripes is consistent with the EBSD orientation information. Such measurements can also be used to identify any additional $180^o$ ferroelectric domains that nucleate across the nonpolar-to-polar phase transition in the form of suppressed SHG intensity at domain walls \cite{Denev2011,Lei2018}. However, there were no such domains observed, despite the presence of ferroelastic domain walls, which are generally expected to be preferential nucleation sites for such features. This observation, while surprising for a polar material, is consistent with the metallic nature of LiOsO$_3$. In insulating ferroelectrics, the depolarizing field created by ordered electric dipoles destabilizes a single domain configuration and drives the formation of $180^o$ polar domains \cite{Lines1977, Guyonnet2014}. Such a depolarizing field would be expected to be screened by free electrons in a metal such as LiOsO$_3$, allowing the single domain configuration to be stable. 

To summarize, the linear and nonlinear optical properties of LiOsO$_3$ were measured, and used to probe its unique polar phase transition. The optical SHG susceptibility was found to be suppressed along [001] owing to a reduction in the Os-O bond anisotropy along the polar axis. Temperature-dependent SHG and Raman measurements were consistent with the occurrence of a continuous order-disorder phase transition at 140 K. Polar ferroelastic domains were observed using SHG microscopy, however no 180$^o$ domain walls were observed.

\section*{Supplementary Material}

The supplementary material contains detailed information on the crystal orientation using electron back-scattering diffraction, spectroscopic ellipsometry, SHG polarimetry, supporting information for the bond anisotropy, and complete Raman spectra over the measured temperature range.

\section*{Acknowledgements}

H.P., Y.Y., V.G., Y.C., and J.C. acknowledge DOE grant DE-SC0012375 for SHG microscopy, sample preparation, and bond anisotropy analysis. H.P., Y.P., and V.G. acknowledge NSF MRSEC CNS grant DMR-1420620 for ellipsometry and Raman studies. D.P. and J.M.R. were supported by ARO (W911NF-15-1-0017). J.C. was supported by the Gordon and Betty Moore Foundation EPiQS Initiative through grant GBMF4534. Y.S. acknowledges National Natural Science Foundation of China grants 11774399, 11474330, and the Chinese Academy of Sciences grants XDB07020100 and QYZDB-SSW-SLH043, and also Kazunari Yamaura in National Institute for Materials Science (NIMS) of Japan for providing high quality LiOsO$_3$ single crystals. L.G. acknowledges NSF grants DMR-0805073, DMR-0958349, DMR-1429428, Office of Naval Research Award No. N00014-06-1-013, and UNF Terry Presidential Professorship. 

H.P. and V.G. acknowledge useful discussions with David Hsieh. H.P. acknowledges Maxwell Wetherington and Julie Anderson for help with sample characterization.

\section*{references}

\bibliography{library}% Produces the bibliography via BibTeX.
\end{document}